\newcount\epsfscale    
\input epsf
\documentstyle[twoside,fleqn,espcrc2]{article}

\newcommand{\AmS}{{\protect\the\textfont2
  A\kern-.1667em\lower.5ex\hbox{M}\kern-.125emS}}

\title{Design of the Muon Collider Lattice: Present Status}
\author{ A. Garren\address{Lawrence Berkeley National Laboratory, Berkeley, CA
94720, USA\protect}\address{Center for Advanced Accelerators, UCLA, 
Los Angeles, CA
90095, USA}, E. Courant$^{\rm c}$, J. Gallardo$^{\rm c}$, R. Palmer\address{Brookhaven National
Laboratory, Upton, New York, 11793, USA}\address{Stanford Linear
Accelerator Laboratory, Stanford, CA 94309, USA}, D. Trbojevic$^{\rm c}$, C.
Johnstone$^{\rm e}$,  K-Y. Ng\address{Fermi National Accelerator Laboratory,  Batavia,
IL 60510, USA}} 
\begin{document}
\begin{abstract}
We discuss a preliminary design for a high luminosity 4 TeV center of mass 
$\mu^+\,\mu^-$ collider ring.
\end{abstract}
\maketitle
\section{INTRODUCTION}
The last component of a muon collider facility, as presently
envisioned\cite{ref00}, is a colliding-beam storage
ring. Design studies on various problems for this ring have  been in progress
over the past year\cite{ref01}. 
In this paper we discuss the current status of the design.

The projected muon currents require very low beta values at the IP,
$\beta^*=3\,{\rm mm},$ in order to achieve the design luminosity of ${\cal L}
=10^{35}\,{\rm cm}^{-2}\,{\rm s}^{-1}.$ The beta values in the final-focus
quadrupoles are roughly $400\,{\rm km}.$ To cancel the corresponding
chromaticities, sextupole schemes for local  correction have been included in
the optics of the experimental insertion\cite{ref02}. The {\it hour-glass}
effect constraints the bunch length to be comparable to $\beta^*.$ To obtain such short
bunches with reasonable rf voltage  requires a very small value of the 
momentum compaction $\alpha,$  which can be obtained by using flexible momentum
compaction (FMC) modules in the arcs\cite{ref03}.

\begin{figure}[tbh]
\epsfxsize=7.0cm \epsfysize=7.0cm \epsfbox{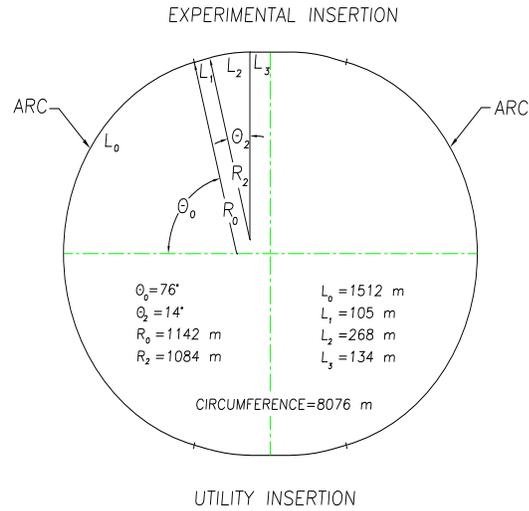}
\caption{The complete collider ring layout.}
\label{fg1}
\end{figure}

A preliminary design of a complete collider ring has now been made; it uses an
experimental insertion and arc modules as described in 
refs.\cite{ref01}-\cite{ref03} as well as a utility insertion. 
The layout of this ring is shown schematically in
Fig.\ref{fg1},
and its parameters are summarized  in Tb.\ref{tb1}. 

\begin{table}[hbt]
\setlength{\tabcolsep}{0.5pc}
\newlength{\digitwidth} \settowidth{\digitwidth}{\rm 0}
\catcode`?=\active \def?{\kern\digitwidth}
\protect \caption{High Energy-High Luminosity $\mu^+\,\mu^-$ Collider.}
\label{tb1}
\begin{tabular}{ll}
\hline
Maximum c-of-m Energy [TeV]& 4 \\
Luminosity ${\cal L}$[$10^{35}$cm$^{-2}$s$^{-1}$] &1.0 \\
Circumference [km] & 8.08\\
Time Between Collisions  [$\mu $s]& 12 \\
Energy Spread $\sigma_e$[units $10^{-3}$] & 2\\
Pulse length $\sigma_z$[mm] & 3\\
Free space at the  IP [m]  & 6.25 \\
Luminosity life time [No.turns] & 900 \\
{\it rms} emittance, $\epsilon_{x,y}^N$ [$10^{-6}$m-rad]& 50.0 \\
{\it rms} emittance, $\epsilon_{x,y}$ [$10^{-6}$m-rad]& 0.0026 \\
Beta Function at  IP, $\beta^*$ [mm]& 3 \\
{\it rms} Beam size at IP [$\mu$m] &2.8 \\
Quadrupole pole fields near IP [T] &6.0 \\
Maximum Beta Function, $\beta_{\rm{max}}$ [km]& 400 \\
Magnet Aperture closest IP [cm] & 12 \\
Beam-Beam tune shift per crossing &0.04 \\
Repetition Rate [Hz]& 15 \\
rf frequency [GHz]& 3 \\
rf voltage [MeV]& 1500 \\
Particles per Bunch [units $10^{12}$]& 2 \\
No. of Bunches of each sign & 2 \\
Peak current ${\cal I}={eNc\over \sqrt{2\pi}\sigma_z}$ [kA]&12.8 \\
Average current ${\cal I}={eNc\over {\rm Circum}}$ [A]& 0.032\\
Horizontal tune $\nu_x$ & 55.79\\
Vertical tune $\nu_y$ & 38.82\\
\hline
\end{tabular} 
\end{table}

Though some engineering features are unrealistic, and the
beam performance needs some improvement, we believe that this study can
serve as the basis for a workable collider design.

The remaining sections of the paper will describe the lattice, show
beam behaviour, and discuss future design studies.
\section{LATTICE}
\subsection{Global structure}
The ring has an
oval shape, with reflection symmetry about 
two perpendicular axes, see Fig.~\ref{fg1}. 
The lattice has two
nearly circular $152^\circ$ arcs 
joined by the experimental and utility insertions. 
Each insertion contains two $14^\circ$ bending sections. 

The two arcs are identical; each contains 22 periodic modules and two
dispersion suppressor modules. The two insertions are geometrically
identical, and each is symmetric about its center. Each half insertion has three 
parts: two straight sections separated by a bending section.
The bending sections are identical in the experimental and utility
insertions, except for sextupole strengths; the 
straight parts have different quadrupole lengths and gradients.
Thus, the focusing structure of the ring has
one superperiod, with reflection symmetry about the line joining the centers
of the two insertions.
 
\subsection{Arc module}
In order to have very
short $3\,{\rm mm}$ bunches in the 2 TeV muon collider, the storage ring 
must be quasi-isochronous, which requires that the momentum compaction
$\alpha$ be very close to zero, where $\alpha$ is defined in terms of
offsets of the momentum $p$ and equilibrium orbit circumference $C$ by
\begin{equation}
\alpha=\frac{\Delta C}{C}/\frac{\Delta p}{p},             \label{I1}
\end{equation}
which may be shown to be equal to
\begin{equation}
\alpha=\frac{1}{C}\oint \frac{D(s)}{\rho(s)}ds\;,            \label{I2}
\end{equation}
where $D$ is the dispersion function, $\rho$ the radius of curvature 
and $s$ is the longitudinal path length
measured along the closed orbit. Since there
is a closed orbit for every value of the momentum, all of these
quantities including $\alpha$ are functions of $p$.

The particle motion 
in longitudinal phase space depends on changes of its arrival time at the 
rf cavities, which depends on changes of circumference and velocity $v$.
To first order the time difference is: 
$\frac{\Delta T}{T_0} =\frac{\Delta C}{C_0} - \frac{\Delta v}{v_0}$
and is related to the change in momentum by 
$\frac{\Delta T}{T_0} = \eta \frac{\Delta p}{p_0} = 
( \alpha_0 - \frac{1}{\gamma^2})
\frac{\Delta p}{p_0},$
where $T_0$ is the time of arrival of the reference particle;  $\Delta T$ and
$\Delta p$\  are the time and momentum deviations, respectively,  of the
off-momentum particle relative to the  synchronous particle with momentum
$p_0$;  $\eta$ is the {\it phase slip} factor; $\gamma$ is the 
relativistic energy, and $\alpha_0 = \alpha(p_0).$
The transition energy $\gamma_t$ is defined by
$\alpha=1/\gamma_t^2$. 
\begin{figure}[tbh]
\epsfxsize=7.0cm \epsfysize=7.0cm \epsfbox{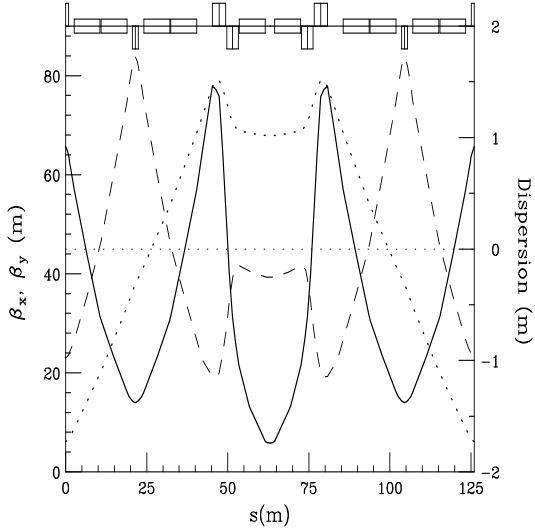}
\caption{Betatron ($\beta_x$ solid-line; $\beta_y$ dash-line)  and 
dispersion (dot-line) functions of an arc-module.}
\label{fg2}
\end{figure}

In an isochronous ring $\eta = 0$, so to first order the arrival time
is independent of the momentum; the condition for which is 
$\gamma_t=\gamma$. For 2 TeV muons $\gamma\approx 2\,10^{4},$ so 
$\alpha\approx 2.5\,10^{-9}.$
In a regular FODO lattice, $\alpha$ is much larger.
To bring the first order value of $\alpha$ to zero requires that
$\langle D/\rho \rangle$ through all of the dipoles be equal to zero. 

In a FODO lattice $\alpha$ is positive. This muon collider ring design
has bending regions in the insertions with a FODO structure whose
contributions to $\alpha$ are positive, so the contributions of the arcs
must be negative with nearly the same magnitude as those of the insertions.
For the present design, the value needed for each arc is 
$\alpha_{arc}=-1.15\,10^{-4}.$

This value of $\alpha_{arc}$ can 
be obtained by building an arc whose periods are FMC modules.
An FMC module\cite{ref03} is a symmetric structure composed
of two FODO cells separated by a matching insertion which
transforms $(\beta_x, \alpha_x, \beta_y, \alpha_y, D, D')$ to 
$(\beta_x, -\alpha_x, \beta_y, -\alpha_y, D, -D').$ 

The contribution to $\alpha$ of the module can be adjusted by choosing the 
appropriate value of $D$ (with $D'=0$) at the end of the module. For the
module design used here (see Fig.\ref{fg2}), the matching insertion contains 
two quadrupole doublets and two dipoles. The two quadrupole gradients and
drift lengths are adjusted to bring $\alpha_x, \alpha_y $ and $D'$
to zero at the center of the module. The number of modules and the 
bending angles of the dipoles are chosen to give the entire arc the
bending angle of $152^\circ$ needed to close the ring.

The arc modules also contain sextupoles; there are two families
adjusted to bring the chromaticities of the arc to zero. 
Alternatively, they could be used to control the quadratic dependence
of $\alpha(p)$, as is discussed in section 4.1.

The parameters of the arc modules are given in Tb.\ref{tb2}.
\begin{table}[hbt]
\setlength{\tabcolsep}{.5pc}
\protect \caption{Arc-module.}
\label{tb2}
\begin{tabular}{lr}
\hline
Total length [m] & 126 \\
Total angle [deg.] & 6.53 \\
No. modules per arc & 22 \\
No. dipoles & 10 \\
No. quadrupoles & 7 \\
No. sextupoles & 3 \\
Dipole length [m] & 8 \\
Dipole field [T] & 9.51 \\
Max. gradient [T/m] & 240 \\
Max. sext. strength [T/m$^2$] & 2074 \\
Tune $\mu_x$ & 0.896 \\
Tune $\mu_y$ & 0.536 \\
Chromaticity $\mu_x'$ & -1.36 \\
Chromaticity $\mu_y'$   & -0.71 \\
Compaction $\alpha$ & $-5.93\,10^{-5}$ \\
Max. $\beta_x$ [m] & 78 \\
Max. $\beta_y$ [m] & 78 \\
Max. $D$ [m] & 1.52 \\
Min. $D$ [m] & -1.73 \\
\hline
\end{tabular} 
\end{table}
\subsection{Dispersion suppressor}
A dispersion suppressor module is located at each arc end, which brings the
dispersion and its slope to zero in the adjacent insertion straight section.

\begin{figure}[tbh]
\epsfxsize=7.0cm \epsfysize=7.0cm \epsfbox{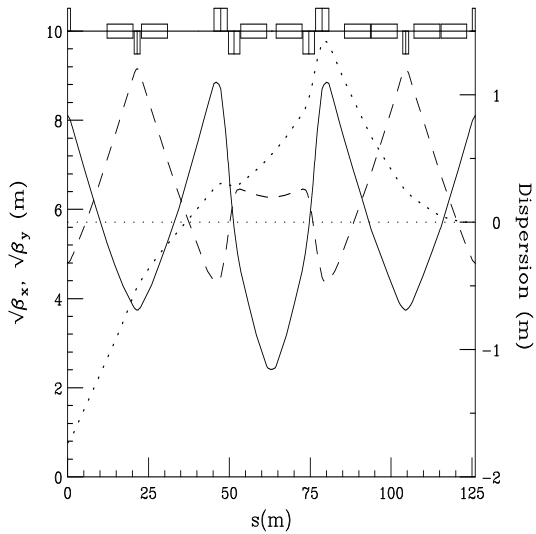}
\caption{Betatron ($\beta_x$ solid-line; $\beta_y$ dash-line)  and 
dispersion (dot-line) functions of a dispersion suppressor module.}
\label{fg3}
\end{figure}

%
The suppressor at the arc-end preceding an insertion  is shown in
Fig.\ref{fg3};  the suppressor following an insertion is obtained by 
reflection. This
suppressor  module is identical to a regular module except that the first four
dipoles have been replaced by two dipoles with normal length and different
field values. 
The lengths, strengths and positions of the quadrupoles and sextupoles are 
the same as in the regular modules.

The suppressor brings the dispersion and its slope to zero at
the end of the arc exactly; however there are small errors in the 
beta functions.
These errors could be removed 
by changing the gradients of some of the quadrupoles.

The parameters of the dispersion suppressor module are given in Tb.\ref{tb3}.

\begin{table}[hbt]
\setlength{\tabcolsep}{.5pc}
\protect \caption{Dispersion suppressor.}
\label{tb3}
\begin{tabular}{lr}
\hline
Total length [m] & 126 \\
Total angle [deg.] & 3.99 \\
No. suppressors per arc & 2 \\
No. normal dipoles & 6 \\
No. special dipoles & 2 \\
No. quadrupoles & 7 \\
No. sextupoles & 3 \\
Dipole length [m] & 8 \\
Field normal dipole [T] & 9.51 \\
Field dipole $\#1$ [T] & 8.72 \\
Field dipole $\#2$ [T] & -7.63 \\
Max. gradient [T/m] & 240 \\
Max. sext. strength [T/m$^2$] & 2074 \\
Tune $\mu_x$ & 0.896 \\
Tune $\mu_y$ & 0.536 \\
Chromaticity $\mu_x'$ & -1.36 \\
Chromaticity $\mu_y'$   & -0.71 \\
Compaction $\alpha$ & $-3.44\,10^{-4}$ \\
Max. $\beta_x$ [m] & 80 \\
Max. $\beta_y$ [m] & 84 \\
Max. $D$ [m] & 1.41 \\
Min. $D$ [m] & -1.73 \\
\hline
\end{tabular} 
\end{table}
\subsection{Experimental insertion} 
\begin{figure}[bht]
\epsfxsize=7.0cm \epsfysize=7.0cm \epsfbox{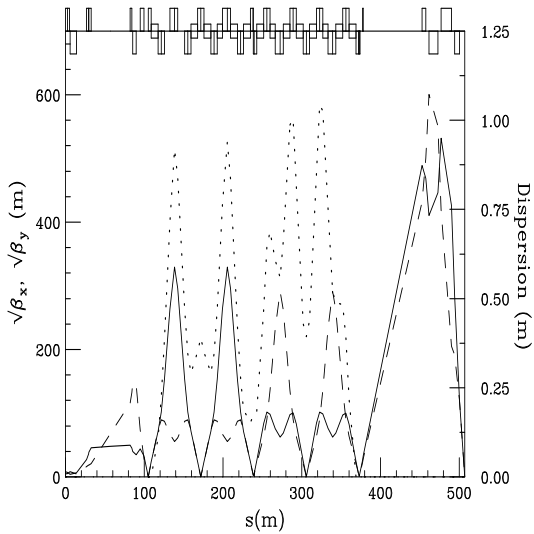}
\caption{Half experimental insertion ($\beta_x$ solid-line; $\beta_y$ dash-line; 
$D$ dot-line).}
\label{fg4}
\end{figure}

 The design of an insertion with 
an extremely low-beta interaction region for a 
muon collider\cite{ref1} presents a challenge 
similar to that encountered in the Next Linear Collider (NLC)\cite{ref2}.
The design used here for each half of the symmetric low-beta insertion 
follows the prescription proposed by Brown\cite{ref5}; 
it consists of two telescopes with a chromatic 
correction section between them. Fig.\ref{fg4} shows the left half of the insertion,
starting at the end of the arc dispersion suppressor and ending at the IP.

The first telescope, called the Matching Telescope (MT), on the left of the
figure, 
brings the beta functions from the arc to a focus of about 3 cm. 
To the right of the MT lies the Chromatic Correction Section (CCS), which 
contains two
pairs of non-interleaved sextupoles. One pair, situated at positions of
maximum $\beta_x$ and large dispersion $D$,  corrects horizontal chromaticity;
the other pair, at maximum $\beta_y$ positions, corrects vertical chromaticity. 
The horizontal-correcting pair is farthest from the IP, 
and the vertical-correcting pair is closest.
The sextupoles of each 
pair are separated by phase advances of $\phi=\pi$  $(\Delta \mu =-0.5),$ 
and they are all located at odd multiples of $\pi /2$ phase intervals from 
the IP.
To the right of the CCS, the Final Focus Telescope (FFT) transports the
beta functions from a focus of a few centimeters to a 3 mm focus at the IP.

The low beta-function values at the IP are obtained with four
strong quadrupoles in the FFT with high beta values; these generate large 
chromaticities which are corrected locally with the two sextupole pairs 
in the CCS. This sextupole
arrangement cancels the second-order geometric aberrations of the sextupoles,
which reduces the second order tune shift by several orders of magnitude. The
momentum bandwidth of the system is limited by third-order aberrations and 
residual second-order amplitude-dependent tune shifts.
These  aberrations arise from: a) small phase errors between the sextupoles and
the final quadruplet; b) finite length of the sextupoles. 

 The residual chromaticities could be reduced with additional
sextupoles at locations with nonzero dispersion, as suggested by
Brinkmann\cite{ref7}. Finally, a system of octupoles could be designed
to correct third-order aberrations. Overall, it is believed 
possible to construct a system with a bandwidth of $\approx 1\,\%.$ 

The most complex part of the insertion is the CCS. A
somewhat oversimplified description follows.
The CCS consists of eight FODO 
cells, each with ${\pi / 2}$  phase advances. The first four cells 
from the left begin
at the center of a QF quadrupole and contain the two horizontal $S_x$ 
sextupoles, which are next to QF's;  the next four
cells begin at the center of a QD and contain the vertical $S_y$ 
sextupoles next to QD's.
The low-beta focus at the beginning of the
CCS repeats itself every two cells and produces the 
high beta values needed in the sextupoles.
The dipoles are placed in a way to cancel the dispersion and its slope at
the ends of the CCS and to produce dispersion maxima near the sextupoles.

The strengths of the sextupoles $S_x$
and $S_y$ are adjusted to minimize the first order chromaticity, while trim
quadrupoles, located at the ends of the fourth and eighth FODO cells, are used to minimize the second order
chromaticity $({\partial^2 \mu / \partial \delta^2}).$ 
The complete insertion has very small residual
chromaticity and is nearly transparent when attached to the arc lattice.

The total length of the half insertion is 507 m;
it contains
44 quadrupoles, 14 sector dipoles and  4 sextupoles. 
A few dipoles have excessive fields and there is no space between many of
the magnets.
\begin{figure}[tbh]
\epsfxsize=7.0cm \epsfysize=7.0cm \epsfbox{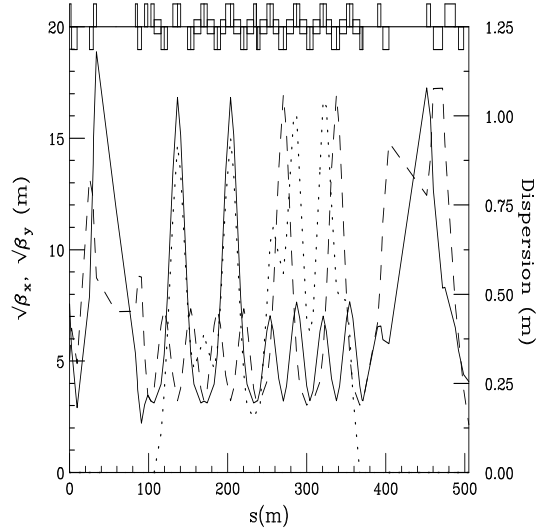}
\caption{Half utility insertion ($\beta_x$ solid-line; $\beta_y$ dash-line; 
$D$ dot-line).}
\label{fg5}
\end{figure}
Parameters of the experimental insertion are given in Tb.\ref{tb4}.
\begin{table}[hbt]
\setlength{\tabcolsep}{.5pc}
\protect \caption{Experimental insertion.}
\label{tb4}
\begin{tabular}{lr}
\hline
Total length [m] & 1014 \\
Total angle [deg.] & 28.3 \\
Mom. compaction $\alpha$ & $1.154\,10^{-4}$ \\
Horizontal tune $\mu_x$ & 6.41 \\
Vertical tune $\mu_y$ & 6.56 \\
Max. $\beta_x$ [km] & 283 \\
Max. $\beta_y$ [km] & 360 \\
Max. dispersion $D$ [m] & 1.04 \\
\hline
\end{tabular} 
\end{table}
\subsection{Utility insertion}
The utility insertion closely resembles the experimental insertion, except 
that the low-beta foci are relaxed in order to lower the 
beta-function maxima by a factor of about 1000, see Fig.\ref{fg5}. 
 This is done by relaxing the
focusing in the two telescopes. The CCS section
is the same as in the experimental insertion, except that the
sextupoles are adjusted to cancel the total chromaticity of the 
utility insertion. Further changes will probably be needed in the future
to better accommodate requirements of injection, RF, and scraping.
%
\section{PERFORMANCE} 
\begin{figure}[tbh]
\epsfxsize=7.0cm \epsfysize=7.0cm \epsfbox{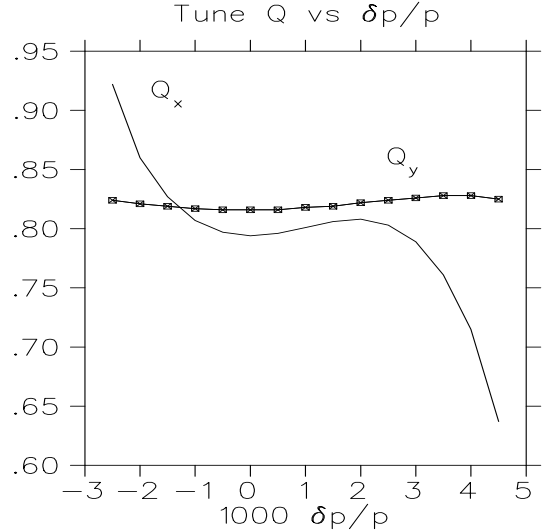}
\caption{Fractional part of the tune shift $Q$ vs  ${\Delta p\over p}.$ } 
\label{fg6}
\end{figure}

\begin{figure}[tbh]
\epsfxsize=7.0cm \epsfysize=7.0cm \epsfbox{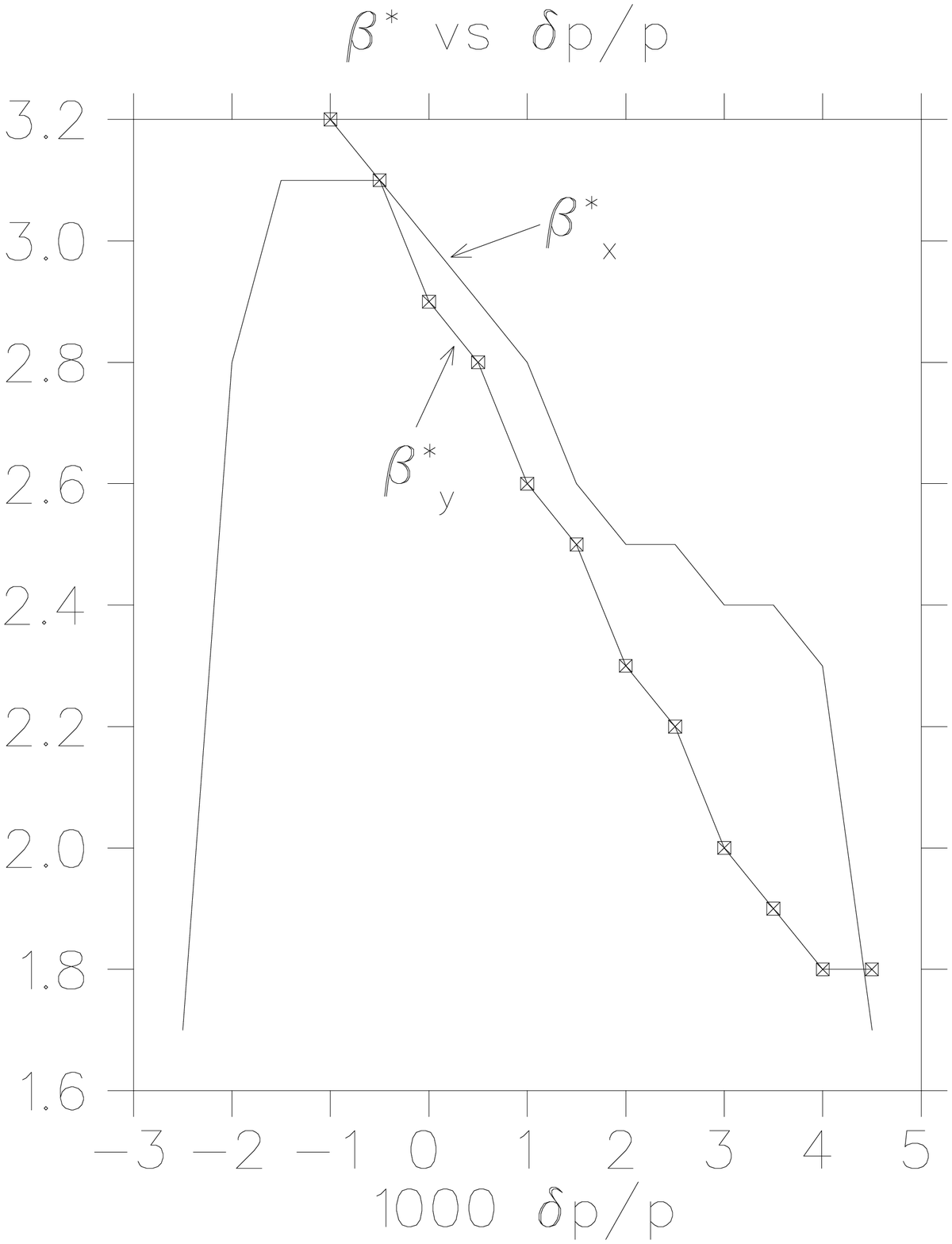}
\caption{Beta function $\beta^*$ vs ${\Delta p\over p}.$}
\label{fg7}
\end{figure}

\begin{figure}[tbh]
\epsfxsize=7.0cm \epsfysize=7.0cm \epsfbox{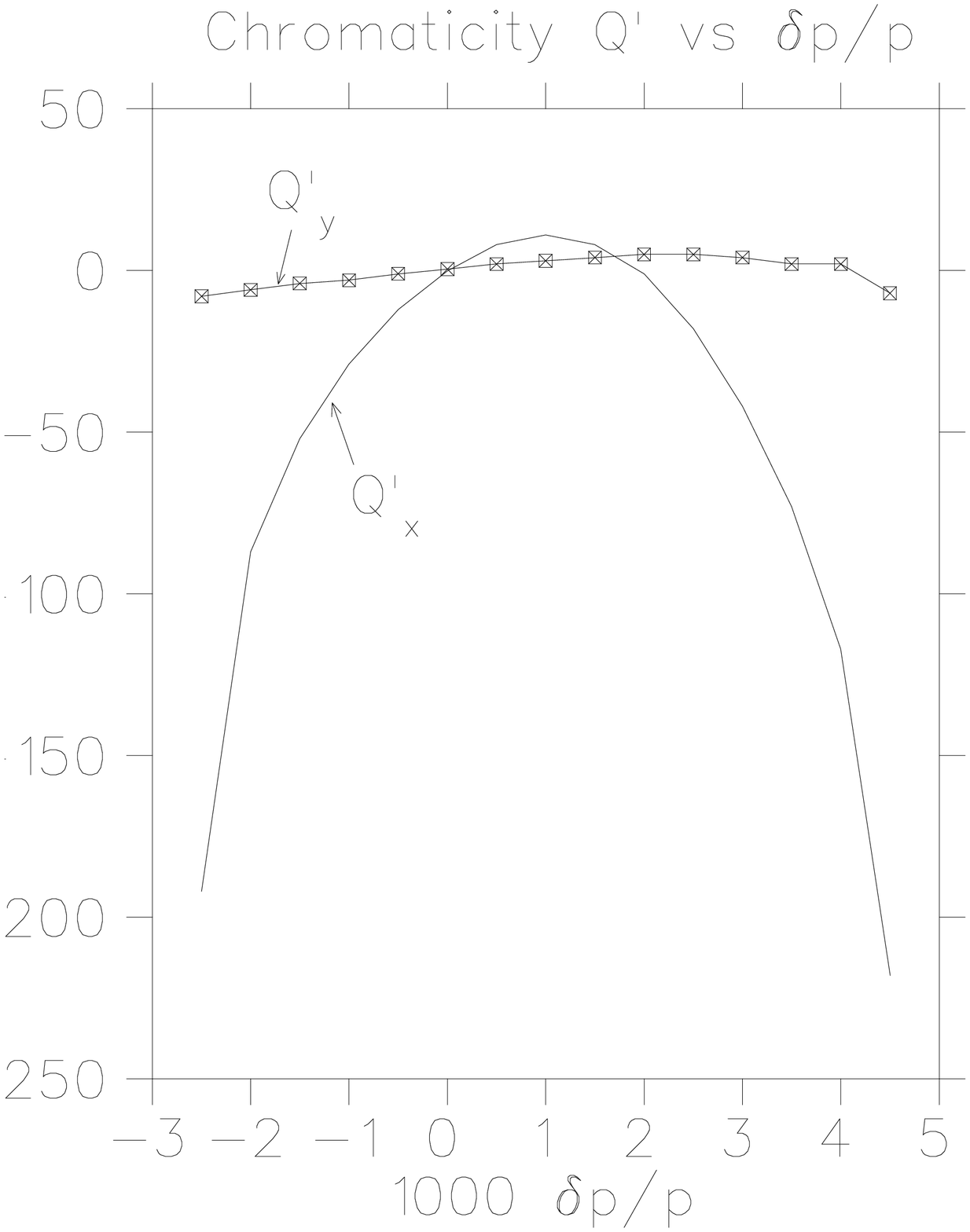}
\caption{Chromaticity $Q'$ vs ${\Delta p\over p}.$ }
\label{fg8}
\end{figure}

\begin{figure}[tbh]
\epsfxsize=7.0cm \epsfysize=7.0cm \epsfbox{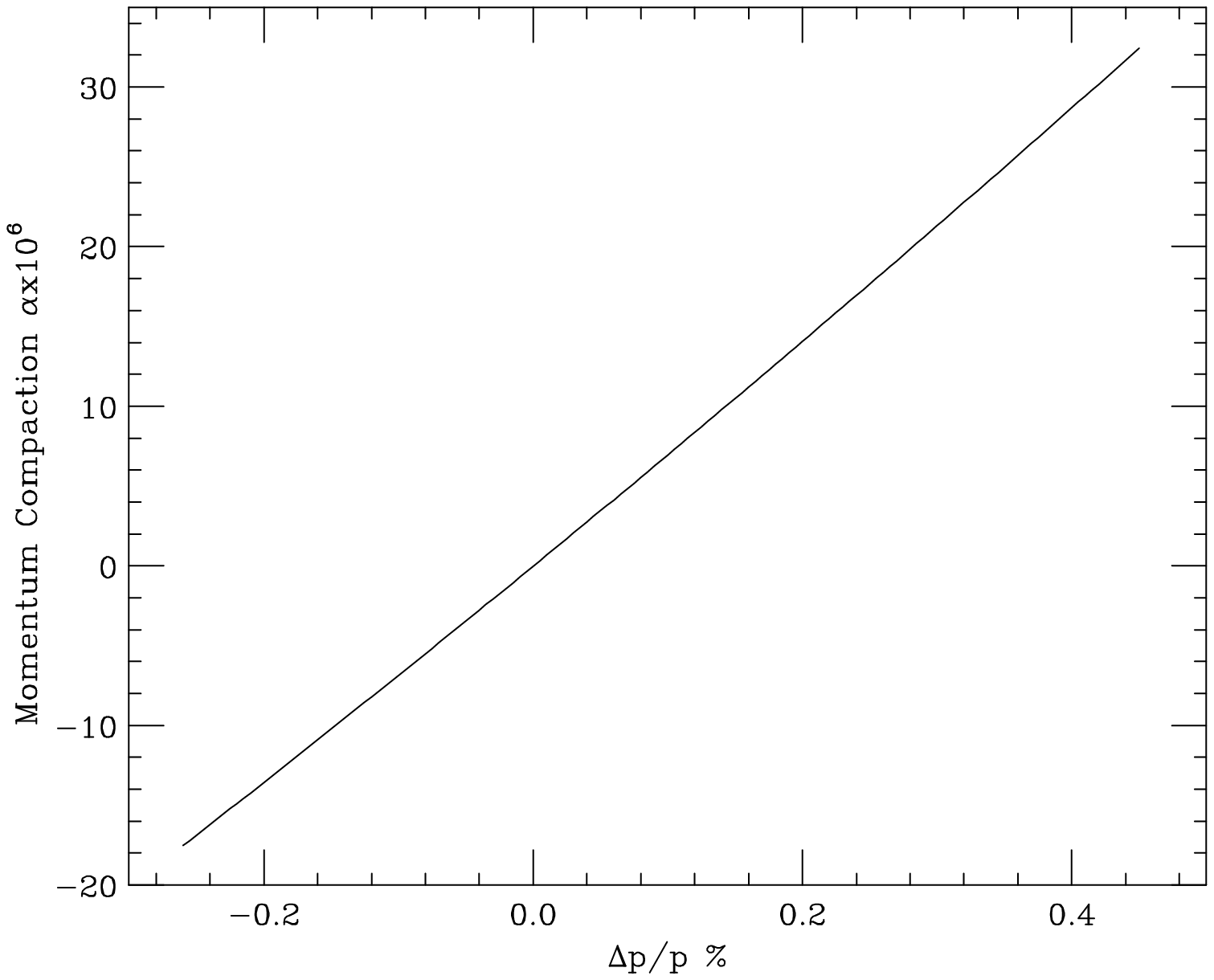}
\caption{Momentum compaction $\alpha$ vs ${\Delta p\over p}.$ }
\label{fg9}
\end{figure}

\begin{figure}[tbh]
\epsfxsize=7.0cm \epsfysize=7.0cm \epsfbox{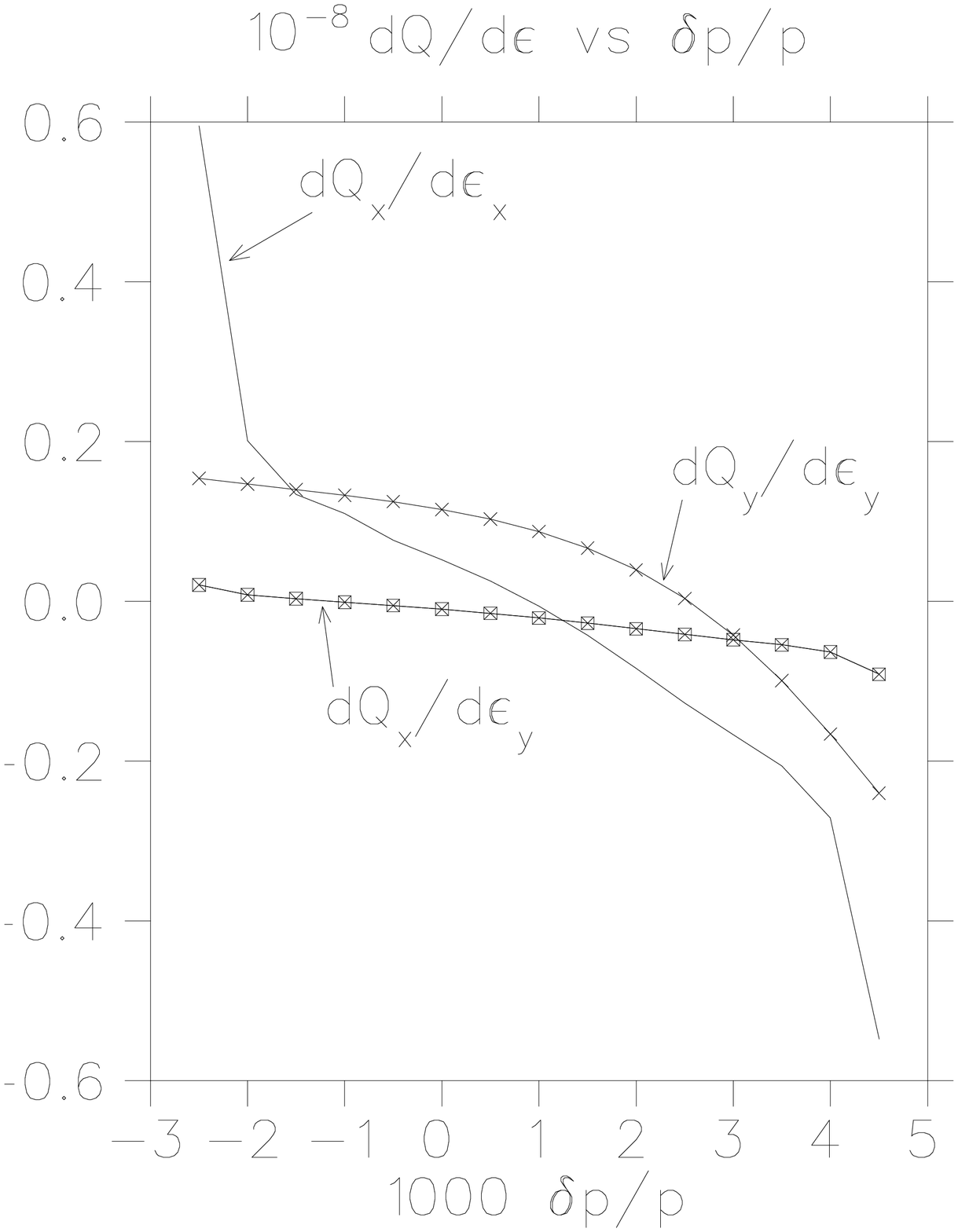}
\caption{Amplitude dependent tune shift ${dQ\over d\epsilon}$ vs ${\Delta
p\over p}.$ }
\label{fg10}
\end{figure}

The variations of the fractional part of the global tunes $Q_x,Q_y$ as functions  
of $\Delta p/p $ are shown in Fig.\ref{fg6}. 
$Q_y$ is essentially flat over a bandwidth of $\pm 0.4\%$, but $Q_x$ has 
non-linear components, although the variation of tune, peak to peak is less
than $0.04$ within a bandwidth of $- 0.15\,\%$ to $0.3\,\%.$ 
The next figures show the momentum dependences of $\beta^*$ (Fig.\ref{fg7}),
chromaticity (Fig.\ref{fg8}), 
momentum compaction $\alpha$ (Fig.\ref{fg9}) and of the amplitude
dependent tune shifts ${dQ / d\epsilon}$ (Fig.\ref{fg10}).

\section{IMPROVEMENTS}
\subsection{Suppression of Spread of Momentum Compaction}
Although the collider ring has been adjusted to be isochronous for the reference
particle, the off-momentum particle will see a small nonzero 
momentum-compaction factor, which is defined to be 
$\alpha(p)=\frac{p}{C}\frac{dC}{dp}.$
 The length of the closed orbit $C$ for an off-momentum particle can be expanded
as a function of momentum offset $\delta=p/p_0-1$, we write 
$C=C_0[1+\alpha_1\delta+\alpha_1\alpha_2\delta^2+{\cal O}(\delta^3)],$ 
 where $\alpha_1\alpha_2$ is considered as a single variable. Consequently, the
momentum compaction seen by the off-momentum particle is 
$\alpha(p)=\alpha_1+(2\alpha_1\alpha_2+\alpha_1-\alpha_1^2)\delta+
{\cal O}(\delta^2).$
 Therefore, for an isochronous ring, there is still a spread of  momentum
compaction $\Delta\alpha=2\alpha_1\alpha_2\delta+{\cal O}(\delta^2)$.  The
first-order  momentum compaction is determined by the dispersion function at
the dipoles. The second-order momentum compaction, however, in addition to
the contribution  from the second-order dispersion function at the dipoles,
contains an extra wiggling term. This wiggling term, equal to one half the
average of the square of the derivative of the dispersion, is a measure of the
additional path length due to the closed orbit wiggling in and out of the reference
orbit. 

\begin{figure}[tbh]
\epsfxsize=6.0cm \epsfysize=7.0cm \epsfbox{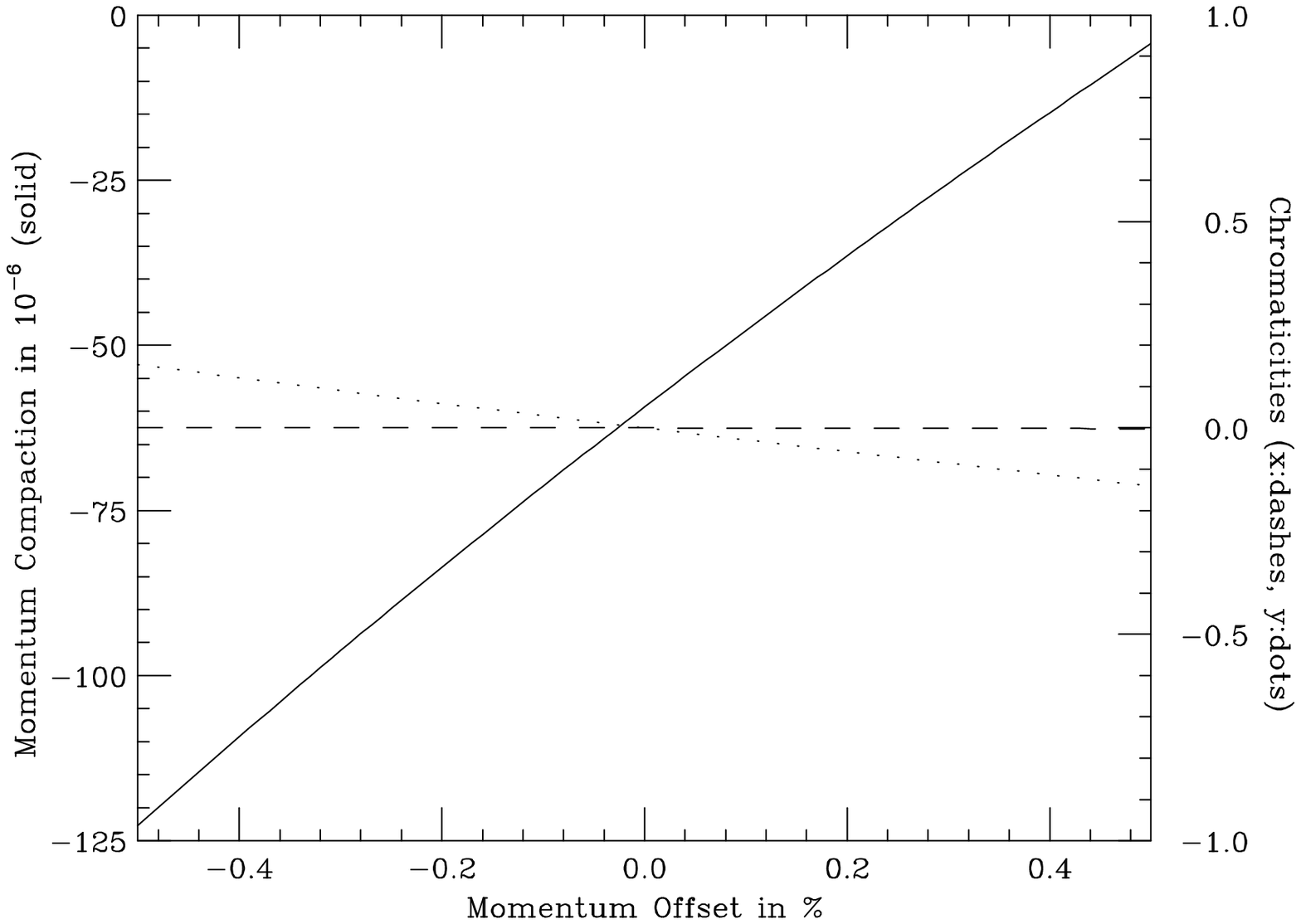}
\caption{Momentum compaction and chromaticities vs ${\Delta p\over p}$ of the
arc-module. }
\label{fg11}
\end{figure}

For normal FODO cells, the dispersion wiggles between 0.65~m and 1.24~m from
the defocusing quadrupole to the focusing quadrupole, for a total of 0.59~m; 
 for a FMC  module, the dispersion
oscillates between  $-1.73$~m and $+1.52$~m, for a total of 3.25~m.  As a
result, the wiggling  term is expected to be  much larger.  This eventually
leads to a much larger $\alpha_1\alpha_2$ than for a FODO cell.  If the
chromatic sextupoles at the IR are used to correct the chromaticities of
the IR, while the sextupoles in arc modules correct chromaticities of the arc
modules, then this isochronous ring has a momentum-compaction factor 
varying almost
linearly with momentum; $\alpha(p)=-4.3334\times10^{-6}$ and
$-122.6758\times10^{-6}$ at $\pm0.5\%$, respectively, for a total spread of
$-188.3424\times10^{-6};$ this is illustrated in Fig.\ref{fg11}. We can also
see that $\alpha_1\alpha_2$ is positive.

In order to have the muons remain bunched, a rather huge rf system
will be necessary for such a large spread of momentum compaction.

\begin{figure}[tbh]
\epsfxsize=7.0cm \epsfysize=7.0cm \epsfbox{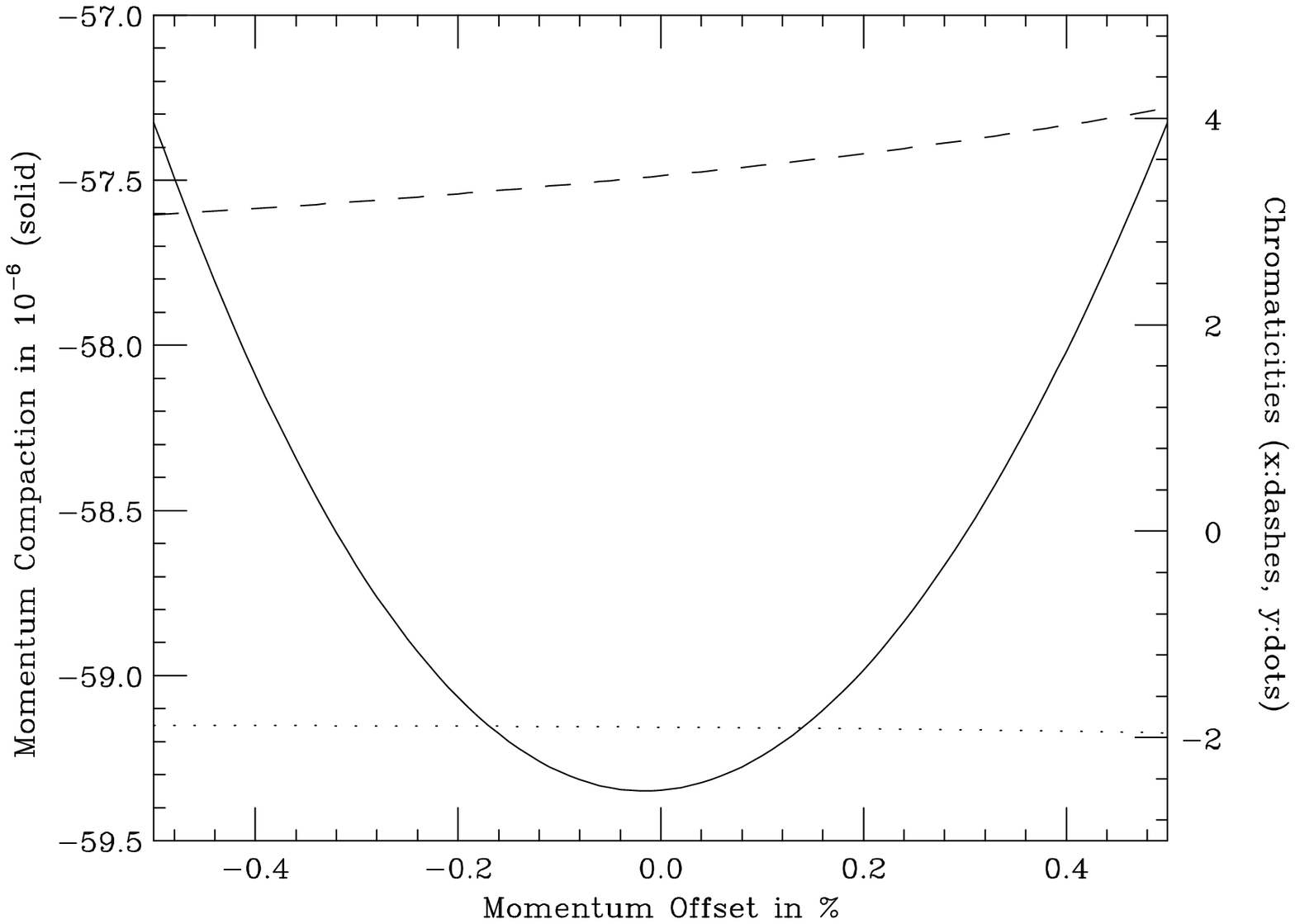}
\caption{$\alpha$ and $\xi$ vs ${\Delta p\over p}$ with 
$\alpha_1\alpha_2$ correction sextupole.  }
\label{fg12}
\end{figure}

Sextupoles affect $\alpha_1\alpha_2$. Consider a
FODO-cell system with sextupoles of strengths $S_F$ and $S_D$, respectively, 
at the focusing and defocusing quadrupoles.  The change in
second-order momentum compaction\cite{ng} is given by 
$\Delta\alpha_1\alpha_2=-(S_F\hat{D}^3+S_D\check{D}^3),$
where $\hat{D}$ and $\check{D}$ are the first-order dispersions at the
focusing and defocusing quadrupoles.  We may expect the behavior to be
similar for our arc modules.  If both sextupoles are used to compensate
for natural chromaticities, the one with $S_F$ will lower $\alpha_1\alpha_2$
while the one with $S_D$ will increase it.  Therefore, we should first
forget chromaticity correction and use only one sextupole of strength $S_F$
on each side of the lower-beta doublets where $\beta_x$ and dispersion is
large.  At this moment, each arc module has been tuned to have
$\alpha_1=-59.3475\times10^{-6}$ in order to maintain isochronocity of the
whole ring.  By adjusting the sextupole strength to the optimum value of
$S_F=0.26623$~m$^{-2}$, the momentum compaction of the arc module becomes
$\alpha(p)=-57.3255\times10^{-6}$ when the momentum offset is $\pm0.5\%$.
The variation of $\alpha(p)$ is plotted in Fig.\ref{fg12} and looks like a
parabola, implying that  $\alpha_1\alpha_2$ has been adjusted to zero and
what is left is the third-order contribution.
  The total swing of
$\alpha(p)$ is now $2.0220\times10^{-6}$.  From the shape of Fig.\ref{fg11},
the momentum-compaction factor of each module should have been adjusted
instead to $\alpha_1=\frac12(-59.3475\times10^{-6}-57.3255\times10^{-6})=
-58.3365\times10^{-6}$, so that eventually the spread of momentum compaction
will become $\pm1.0110\times10^{-6}$, which is acceptable for a
moderate-size rf system.

The introduction of this $\alpha_1\alpha_2$-correction sextupoles bring
the chromaticities (here we denote chromaticity with the symbol $\xi$) of each arc module to $\xi_x=+3.125$ to $+4.099$ and 
$\xi_y=-1.955$ to $-1.884$ when the momentum offset varies from $-0.5\%$ 
t0 $+0.5\%$, as illustrated in Fig.\ref{fg12}.
There are 40 arc modules and they contribute therefore chromaticities
of $\xi_x=+140$ and $\xi_y=-72$ to the whole collider ring.  But
these chromaticities are only very tiny compared with the $-6000$ units
from the IR, and can be removed by making minor adjustment to
the chromatic correction sextupoles of the IR.  There will
still be a spread of chromaticities as a function of momentum due to 
the arc modules, with a total of $\Delta\xi_x=39$ and $\Delta\xi_y=2.84$.

\begin{figure}[tbh]
\epsfxsize=7.0cm \epsfysize=7.0cm \epsfbox{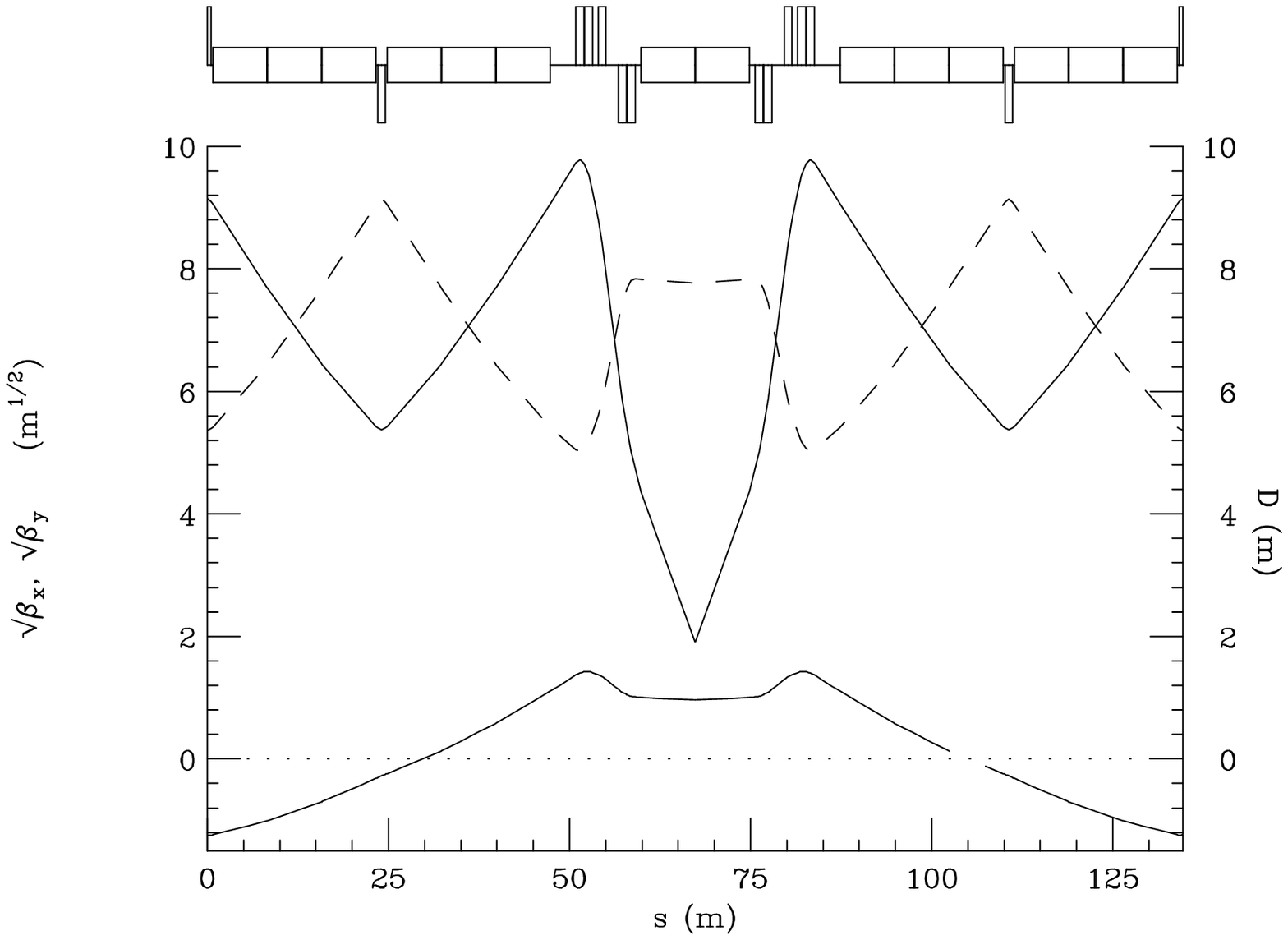}
\caption{Proposed arc module.  }
\label{fg13}
\end{figure}

If the spread of momentum compaction and the spreads of chromaticities
are still too large, we can 
construct another similar arc module that has a smaller dispersion
wiggling.  We have a design in Fig.\ref{fg13} that employs quadrupoles 
with strengths reduced roughly by 1/3  in the FODO regions, but
almost doubled in the low-beta region. 

\begin{figure}[tbh]
\epsfxsize=7.0cm \epsfysize=7.0cm \epsfbox{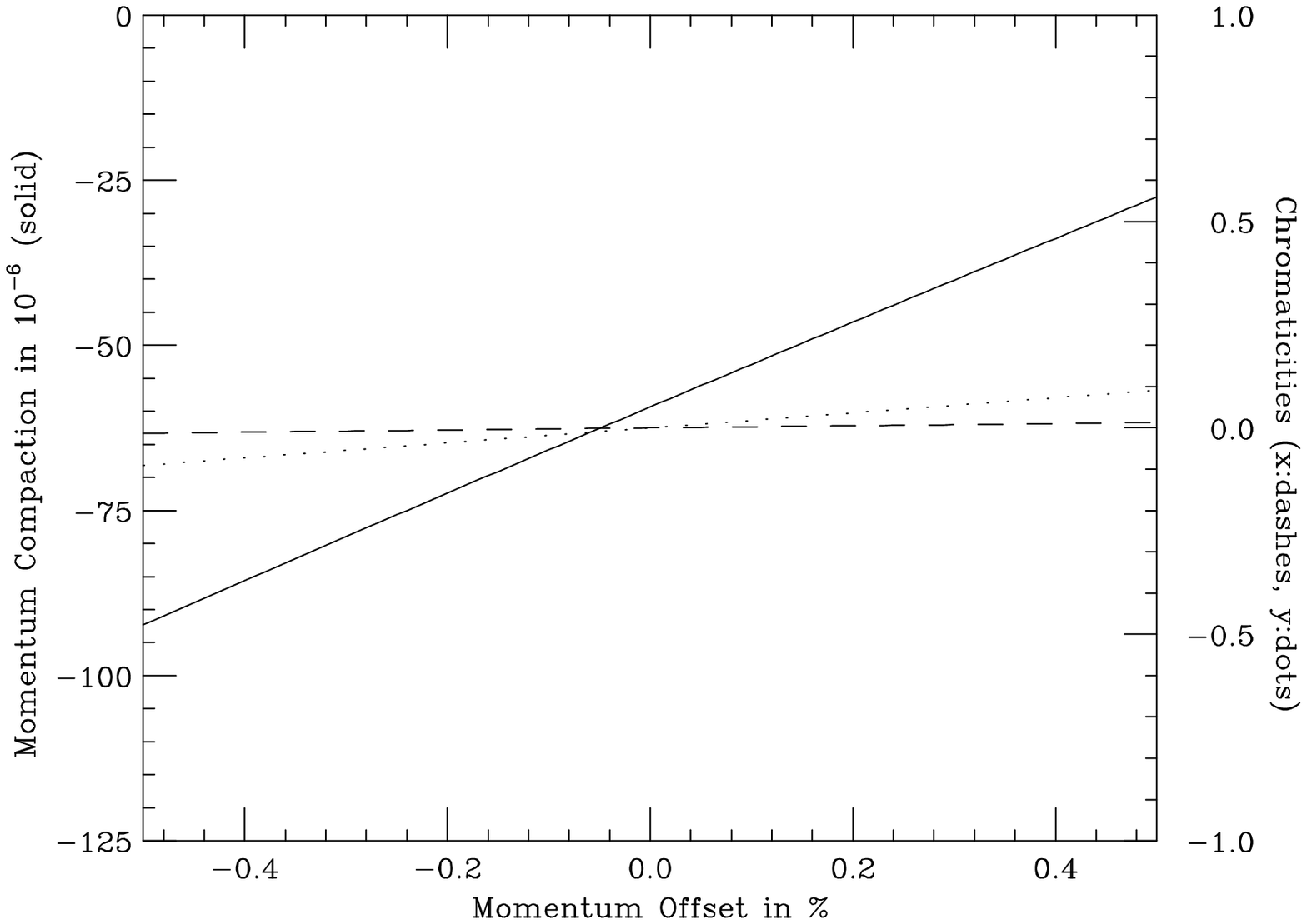}
\caption{Momentum compaction and chromaticities vs ${\Delta p\over p}.$ }
\label{fg14}
\end{figure}

 This module has a maximum and
minimum dispersion of $+1.4228$~m and $-1.2514$~m.  When its chromaticities
are corrected to zero with focusing and defocusing  sextupoles, it gives a 
total spread of $\alpha(p)$
equal to only $64.7327\times10^{-6}$ when the momentum offset is
$\pm0.5\%$, as shown in Fig.\ref{fg14} drawn in the same scale as
Fig.\ref{fg12}.
  
If only a pair of sextupoles $S_F$
is used to compensate for $\alpha_1\alpha_2$ as in above, 
at the optimum sextupole strength of $S_F=0.208081$~m$^{-2}$,
the spread of momentum compaction
will only be $\pm0.3990\times10^{-6}$,
which is about 2.5 times less than the former arc module.
\begin{figure}[bht]
\epsfxsize=6.5cm \epsfysize=7.0cm \epsfbox{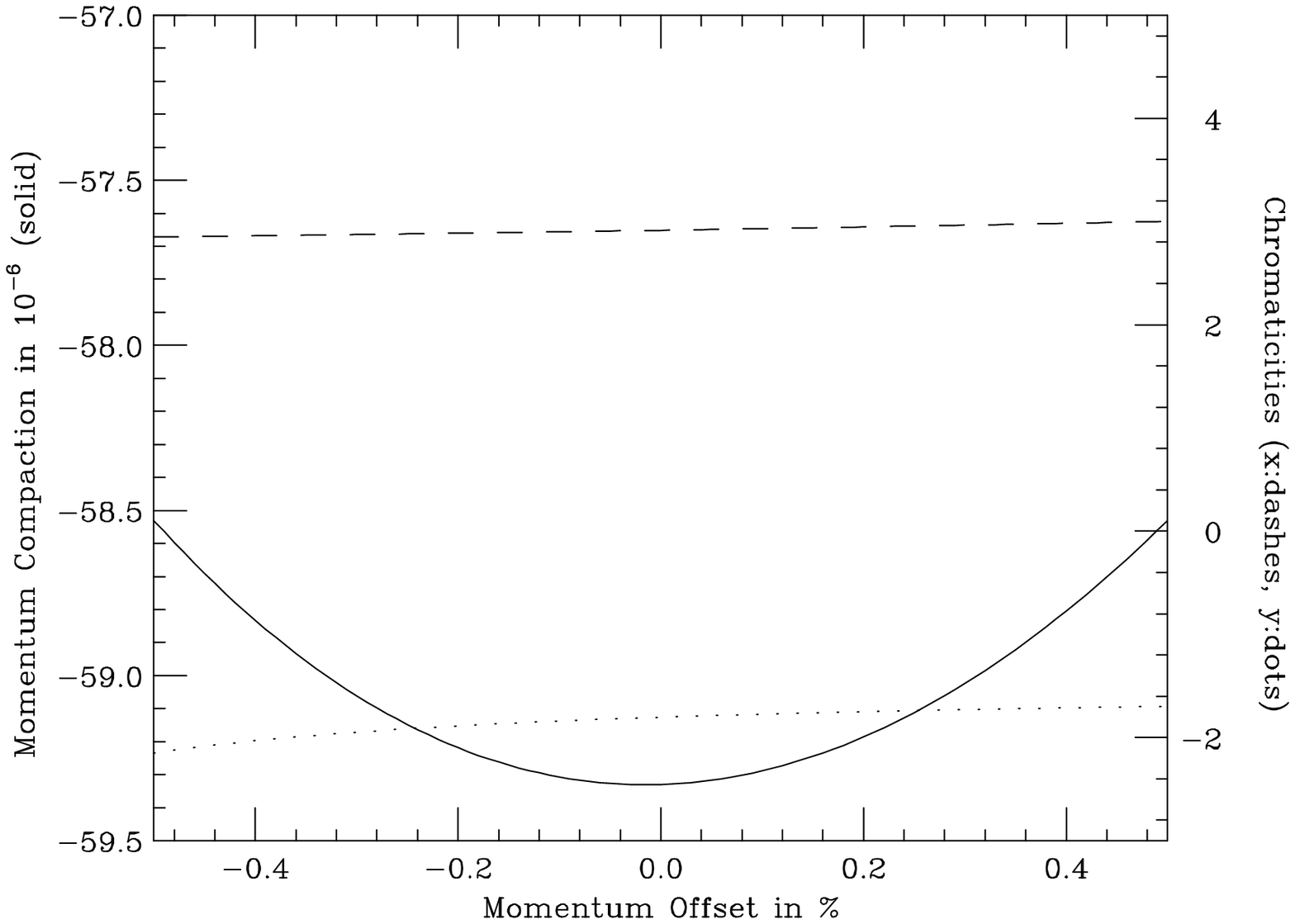}
\caption{$\alpha$ and $\xi$ vs ${\Delta p\over p}$ with $\alpha_1\alpha_2$ 
correction sextupole.  }
\label{fg15}
\end{figure}
Meanwhile, the chromaticities introduced are $\xi_x=+3.002$ to $+2.849$ and 
$\xi_y=-2.149$ to $-1.698$.  Both the chromaticities and their spreads are much
smaller than the arc modules used in the present design.  Fig.\ref{fg15} shows
the momentum compaction and chromaticities for this module as a comparison with
Fig.\ref{fg12}.
\section{CONCLUSIONS}
A possible lattice  for a muon collider has been described. The design
satisfies most of the collider requirements, although it is not fully
realistic.
Error and tolerance analyses are yet to be performed as well as tracking
to determine the dynamic aperture and achievable luminosity. 

In order to make the final-focus design realistic, drift spaces must be
introduced between all magnet elements, and  the lengths of the insertions will
have to be increased in order to achieve the required dispersion values in the
sextupoles with reasonable dipole fields.

\section*{ACKNOWLEDGMENTS}
This research was supported by the U.S. Department of Energy  under Contract
No. DE-ACO2-76-CH00016. (RBP) gratefully acknowledge stimulating discussions
with J. Irwin, O. Napoly and K. Oide.



\begin{thebibliography}{99}
\bibitem{ref00}R. Palmer, et al. {\it Muon Collider Design}, this Proceedings.
\bibitem{ref01}N.M. Gelfand, {\it A Prototype Lattice Design for a 2 TeV
$\mu^+\mu^-$ Collider}, Fermilab Report TM-1933, 1995; King-Yuen Ng,
presentation at the {\it 9th Advanced ICFA Beam Dynamics Workshop: Beam
Dynamics and Technology Issues for $\mu^+\mu^-$ Colliders}, Montauk, New York,
Oct 15-20, 1995, to be published; D. Trbojevic, et al., {\it Design  of the
Muon Collider Isochronous Storage Ring Lattice}, submitted to the proceedings
of the Micro Bunches Workshop, BNL, Sep. 1995, to be published; C. Johnstone
and K.-Y. Ng, {\it Interaction Regions for a Muon Collider },  submitted to the
proceedings of the Micro Bunches Workshop, BNL, Sep. 1995; to be published.
\bibitem{ref02}Juan C. Gallardo and Robert B. Palmer, {\it Final Focus System
for a Muon Collider: A Test Model}, this Proceedings.
\bibitem{ref03}S. Y. Lee, K. Y. Ng and D. Trbojevic, {\it 
Minimizing dispersion in flexible-momentum-compaction lattices}, 
Phys. Rev. E {\bf 48}, 3040 (1993)
\bibitem{ref1}R. B. Palmer, {\it Beam Dynamics in a Muon Collider} Beam
Dynamics Newsletter, No.8 (1994) 27, Eds. K. Hirata, S. Y. Lee and F. Willeke.
\bibitem{ref2} F. Zimmermann, et al., {\it A Final Focus System for the Next
Linear Collider}, SLAC-PUB-95-6789, June 1995, presented at PAC95, Dallas,
Texas, May 1-5, 1995; to be published; O. Napoly, {\it CLIC Final Focus System:
Upgraded version with increased bandwidth and error analysis}, DAPNIA/SEA 94
10, CLIC Note 227, 1994. \bibitem{ref5}K. Brown, {\it A conceptual Design of
Final Focus Systems for Linear Colliders}, SLAC-PUB-4159, (1987).
\bibitem{ref7}R. Brinkmann, {\it Optimization of a Final Focus System for Large
Momentum Bandwidth}, DESY-M-90-14, 1990.
\bibitem{johnsen}K.\ Johnsen, {\it Effects of Nonlinearities on Phase
Transition}, Proc. CERN Symposium on High Energy Accelerators and
Pion Physics, Geneva, Vol.{\bf 1}, 106 (1956).
\bibitem{ng}K.Y.\ Ng, ``Higher-Order Momentum Compaction for a
Simplified FODO Lattice and Comparison with SYNCH,''
Fermilab Internal Report FN-578, 1991.
\end{thebibliography}
\end{document}